\documentclass{article}%
\usepackage{amsmath}
\usepackage{amsfonts}
\usepackage{amssymb}
\usepackage{graphicx}%
\newtheorem{theorem}{Theorem}

\newtheorem{definition}[theorem]{Definition}

\newtheorem{notation}[theorem]{Notation}

\newtheorem{proposition}[theorem]{Proposition}
\newtheorem{remark}[theorem]{Remark}

\ifx\pdfoutput\relax\let\pdfoutput=\undefined\fi
\newcount\msipdfoutput
\ifx\pdfoutput\undefined\else
\ifcase\pdfoutput\else
\ifx\paperwidth\undefined\else
\ifdim\paperheight=0pt\relax\else\pdfpageheight\paperheight\fi
\ifdim\paperwidth=0pt\relax\else\pdfpagewidth\paperwidth\fi
\fi\fi\fi

\begin{document}

\title{Finding the Rikitake's attractors by parameter switching}
\author{\emph{This paper is dedicated to the memory of Professor Adelina Georgescu} \mbox{}
\and Marius-F. Danca$^{1,2}$,~Steliana Codreanu$^{3}$\\$^{1}$Department of Mathematics and Computer Science, Avram Iancu University, \\Cluj-Napoca, Romania, \\$^{2}$Romanian Institute of Science and Technology, Cluj-Napoca, Romania\\$^{3}$Department of Theoretical Physics and Computation, Babes-Bolyai University,\\Cluj-Napoca, Romania,}
\maketitle

\begin{abstract}
In this paper the attractors synthesis algorithm for a class of dissipative
dynamical systems with hyperbolic equilibria, presented in \cite{Danca et al},
is applied to generate any attractor of the Rikitake system. By switching
periodically, or even randomly, the control parameter inside a given set of
values, during any finite time interval while the attractor is numerically
approximated, any attractor can be generated. Beside the extension of the
synthesis algorithm to systems with non-hyperbolic equilibria, we have found
for the Rikitake system, a new intriguing transient which, occurring for a
long time interval, is difficult to be numerically found due to the known
system instability along the $x_{3}$-axis.

\end{abstract}


\section{Introduction}

The paleomagnetic records of the Earth's magnetic field show that the field
has changed its polarity many times along geological history (hundreds of
times during the last 160 million years). But intervals among such geomagnetic
polarity reversals are highly irregular. Thus while their average is about
$7.10^{5}$ years, there are intervals as long as $3.10^{7}$ years without
polarity change, but with large deviations of the poles from actual positions.

\begin{figure}
\begin{center}
  \includegraphics[clip,width=0.5\textwidth]{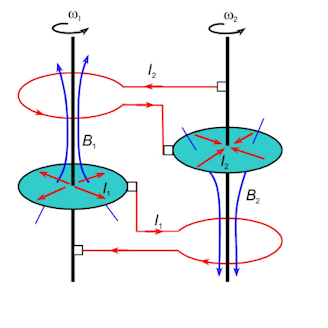}
  \caption{Rikitake dynamo (sketch).}\label{fig1}
  \end{center}
\end{figure}

There are many factors which can affect the Earth's magnetic field.
For example the Reynolds number of the Earth's liquid core is
believed to be of the order of $10^{8}$, i.e. sufficiently large
that the flow of electrical currents in the liquid core be
turbulent. Hence the pattern of the Earth's magnetic field is very
complex (see for ex.\cite{Glatzmaier,Glatzmaier2}). Because of such
great complexity, to study the reversals of the Earth's magnetic
field, relatively simple mechanical dynamos had been proposed as
analog models \cite{Rikitake, Chillingworth, Hide}. One of them is
the Rikitake dynamo of two frictionless coupled disks, which is a
paradigm of the geomagnetic field behavior, proposed by the Japanese
geophysicist Rikitake \cite{Rikitake}. This model can be considered
only as a special case for the real geodynamo, which is obviously a
high order physical system with very large degrees of freedom
\cite{Buffet,Hoyng}. \smallskip The Rikitake dynamo is composed of
two conducting rotating disks which are connected to two coils so
that the current in each coil feeds the magnetic field of the other
(see Fig.\ref{fig1}). Each circuit has the same self-inductance $L$
and electrical resistance $R$, and for each disk dynamo a constant
mechanical torque $G$ is applied from outside on the axis, so that
it can rotate with the angular velocity $\omega.$

The currents $I_{1}$ and $I_{2}$ in the circuits and the related voltages
$V_{1}$ and $V_{2}$ are connected by the well known relations for the $R,~L$ circuits%

\[%
\begin{array}
[c]{c}%
RI_{1}+L\frac{dI_{1}}{dt}=MI_{1}\omega_{1},\\
RI_{2}+L\frac{dI_{2}}{dt}=MI_{2}\omega_{2},
\end{array}
\]

\noindent where $MI_{1}\omega_{1}$ and $MI_{2}\omega_{2}~$are the voltages
$V_{1}$ and $V_{2}$ , and $M$ is the mutual inductance. \smallskip If $C$ is
considered the moment of inertia of each disk, these relations can be rescaled
and lead finally (see for ex \cite{Cook}) to the model equations%

\begin{equation}%
\begin{array}
[c]{cl}%
\overset{.}{x}_{1}= & x_{2}\ast x_{3}-ax_{1},\\
\overset{.}{x}_{2}= & (x_{3}-p)x_{1}-ax_{2},\\
\overset{.}{x}_{3}= & 1-x_{1}x_{2},
\end{array}
\label{IVP}%
\end{equation}

\noindent\noindent with $a=R\sqrt{\frac{LC}{GM}}$, and $\ p=\left(  \omega
_{1}-\omega_{2}\right)  \sqrt{\frac{CM}{GL}}$. The control parameter $p$ is
considered to be positive.

The system is invariant under the change $\left(  x_{1},x_{2},x_{3}\right)
\longmapsto\left(  -x_{1},-x_{2},x_{3}\right)  $. Therefore if $\left(
x_{1}(t),x_{2}(t),x_{3}(t)\right)  $ is a solution to (\ref{IVP}), then
$\left(  -x_{1}(t),-x_{2}(t),x_{3}(t)\right)  $ is also a solution. \smallskip

The system is dissipative the divergence being negative $divf(x)=-2p,~$where
the vector function $f:\mathbb{R}^{3}\rightarrow\mathbb{R}^{3}$ is the right
hand side of (\ref{IVP}).

The equilibrium points are

\[
X_{1,2}\left(  \pm k,\pm\frac{1}{k},ak_{{}}^{2}\right)  ,\text{ with~~}k_{{}%
}^{2}=\frac{1}{2a}\left(  p+\sqrt{p^{2}+4a^{2}}\right)  ,
\]

\noindent and they are not hyperbolic.

The Rikitake model is still intensely investigated, not only for its physical
interest, but especially for the richness of its dynamical behavior (see for
example \cite{Jaume Llibre,Plunian et al}). In our numerical research of the
dynamics of this system, we have found an interesting 'transient attractor'
(TA) persisting for a long time interval before the trajectory reaches one
attractor. \ The trajectory generating this transient passes through the
'real' attractor several times. TA size is disproportionately high compared to
the size of real attractor. Moreover, it is interesting to see that TA is very
unstable being related to the $x_{3}$-axis.

For that reason our paper is focused on two directions: primarily, to present
TA with its proper behavior and secondarily to synthesize any Rikitake
attractor by using the attractor synthesis algorithm (introduced in
\cite{Danca et al}), that switches the control parameter of the system for
finite time intervals, while the model is numerically integrated. Therefore,
this algorithm firstly applied to systems with hyperbolic equilibria, is
extended here to another class of systems. One of the main algorithm benefits
is the fact that it allows the generation of any attractors, because of the
convexity property presented in Section 2, even if for some objective reasons,
some parameter values are not accessible.

The organization of the paper is as follows: Section 2 describes the synthesis
algorithm while in Section 3 the TA is presented beside the application of the
synthesis algorithm. The Conclusion summarizes the results of this work.

\smallskip

\smallskip

\section{\smallskip Synthesis algorithm}

Let us consider the following Initial Value Problem (IVP)%

\begin{equation}
\dot{x}(t)=f(x(t))+p\left(  t\right)  Ax(t),\qquad x(0)=x_{0},\qquad t\in
I=[0,\infty), \label{eq1}%
\end{equation}

\noindent where $f:\mathbb{R}^{n}\rightarrow\mathbb{R}^{n}$ is a nonlinear
vector function, $x_{0}\in\mathbb{R}^{n}~$and $p:I\rightarrow\mathbb{R}^{n}$
is a piece-wise continuous periodic function with period $T~$and mean value
$p^{\ast}$, i.e.%

\begin{equation}
\frac{1}{T}\int_{t}^{t+T}p(u)du=p^{\ast},~~~~t\in I, \label{eq2}%
\end{equation}

\noindent and $A$ is a real $n\times n$ squared matrix.

Next, we make the following assumption

\textbf{H1 }The IVP (\ref{eq1}) admits unique solutions.

In \cite{Yu} it is proved that the solutions of the IVP (\ref{eq1}) for the
case of continuous systems with respect to the state variable, and of the
corresponding \emph{averaged model}, expressed as follows%
\begin{equation}
\dot{x}(t)=f(x(t))+p^{\ast}Ax(t),\qquad x(0)=x_{0},~~~~~t\in I, \label{eq3}%
\end{equation}

\noindent may have arbitrarily close solutions. This means that switching $p$
in the IVP (\ref{eq1}) following some periodic scheme within a selected set of
values, while the IVP is numerically integrated, the solutions remain close
enough to the solutions of the IVP (\ref{eq3}). The analysis is carried out
based on the averaging theory \cite{Sanders}.

The class of systems modeled by (\ref{eq1}) includes known dynamical systems
such as: Lorenz, R\"{o}ssler, Chen, Lotka-Volterra, L\"{u}, minimal networks,
neuronal networks, a class of lasers, etc. Moreover, by using several
computational tests, we verified that the switching algorithm applies not only
to continuous dynamical systems, but also to fractional-order
systems\footnote{For example, the switching algorithm was applied successfully
to a fractional variant of L\"{u} system \cite{Kai}, whose chaotic behavior is
analyzed in \cite{Chen}.}.

While in the mentioned examples we have studied systems with hyperbolic
equilibrium points, in this paper we prove numerically and computationally,
that the algorithm can be applied to the Rikitake's system which has
non-hyperbolic equilibria.

Despite the fact that could be some differences between computation and
theory, the numerical integration of (\ref{eq1}) can generally give excellent
approximations to the trajectories within the invariant sets. The trajectories
that start near an attractor will stay near and they will be shadowed by
orbits within the attractor because attractors arise as the limiting behavior
of trajectories. Therefore, the shadowing property \cite{Coombes} enables us
to recover long time approximation properties of numerical trajectories
necessary in our numerical computations.

Roughly speaking, a \emph{global attractor} can be viewed as a region of a
dynamical system's state space where some of the system trajectories can enter
and not leave, and which contains no smaller such region \cite{Kapitan}. A
global attractor contains all the dynamics evolving from all possible initial
conditions. In other words, it contains all the solutions, including the
stationary and periodic ones, as well as chaotic attractors, relevant to the
asymptotic behaviors of the system.

The term \emph{local attractor} is sometimes used to denominate non-global
attractors (see e.g. \cite{Hirsch}). The global attractors may contain several
local attractors. Therefore, a global attractor may be considered to be
composed of a set of all local attractors. Each of them only attract
trajectories from a subset of initial conditions, specified by its basin of attraction.

\begin{remark}
\label{rem cu attr}For the sake of simplicity, but without loss of generality,
in this paper when a global attractor is composed by several local attractors,
only one of the single local attractors will be considered.
\end{remark}

The attractors are numerically approximated using some scheme for ODEs with
fixed step size $h,$ after the transients are neglected (see e.g. \cite{Foias}).

We have shown via numerical approach and computer simulations, that while the
model (\ref{eq1}) is numerically integrated, if one switch the control
parameter at finite time intervals, the obtained approximated attractor
(\emph{synthesized attractor}) is approximately identical to the one
corresponding to $p=p^{\ast}$ (\emph{averaged attractor}) for whatever
considered set of values for $p.$ The algorithm, we call hereafter
\emph{synthesis algorithm }(SA), consists in using a time-periodically
parameter switching, according to some designed rule. It will be demonstrated
empirically, that any desired attractor can be duly obtained by the proposed
switching scheme. Moreover, we found out numerically and computationally that
SA can be applied not only via some periodic parameter switching rule (as
analytically proved in \cite{Yu}), but using any kind of random switching way.

\begin{notation}
Let $\mathcal{A}$ be the set of all attractors depending on the parameter $p$,
including attractive stable fixed points, stable limit cycles and chaotic
attractors; let $\mathcal{P\subset}\mathbb{R}$ be the set of the admissible
values of $p$~and $\mathcal{P}_{N}=\{p_{1},p_{2},\ldots,p_{N}\}\subset$
$\mathcal{P}\,$ a finite ordered subset of $\mathcal{P}$ which determines the
set of attractors $\mathcal{A}_{N}=\{A_{p_{1}},A_{p_{2}},\ldots,A_{p_{N}%
}\}\subset\,\mathcal{A}$.
\end{notation}

\begin{remark}
It is natural to introduce a bijection between the set of all admissible
values of $p$ and the set of all attractors\footnote{The rigourous proof of
this afirmation, remains a future objective.}. Therefore, giving any $p$, a
unique attractor is specified, and vice versa. Also, via this bijection, the
order over $\mathcal{P}$ induces an order over $\mathcal{A}.$\newline
\end{remark}

For the sake of simplicity, unless necessary, we denote the attractors
$A_{p_{i}}$ simply by $A_{i}.$

We assume that we can access all the values of $\mathcal{P}_{N}=\{p_{1}%
,p_{2},\ldots,p_{N}\}~$for which the system behaves stably or chaotically.

With a chosen finite subset $\mathcal{P}_{N}$, the SA relies on the
following deterministic time switching rule applied indefinitely on
$I,$ while a numerical method with fixed step size $h$ integrates
the IVP (\ref{eq1})

\begin{equation}
\lbrack p_{1}|_{I_{1}},p_{2}|_{I_{2},}\ldots,p_{N}|_{I_{N}}],\text{ }p_{i}%
\in\mathcal{P}_{N},\text{ }i=1,2,\ldots,N, \label{timp}%
\end{equation}

\noindent where $I_{i}$,$~i\in\{1,\ldots,N\}$ are finite consecutive (adjoint)
time subintervals of length $\Delta t_{i}$, for $i=1,2,\ldots,N.$ (\ref{timp})
means that in each interval $I_{i},$ $p=p_{i},~i=1,2,\ldots,N.$ In other
words, $p$ is a piece-wise continuous (constant) function $p:I_{i}%
\mathbb{\longrightarrow\mathcal{P}}_{N},~~p(t)=p_{i},~~$for $t\in
I_{i},~~i=1,2,\ldots,N.$

The simplest way to implement numerically (\ref{timp}) can be
described by the scheme
\begin{equation}
\lbrack p_{1}|_{m_{1}h},~p_{2}|_{m_{2}h},\ldots,p_{N}|_{m_{N}h}],
\label{scheme}
\end{equation}

\noindent where $\Delta t_{i}$ is chosen to have the length $\Delta
t_{i}=m_{i}h$ with $m_{i}~$positive integers (see Section 3). SA acts as
follows: in the first time subinterval $I_{1}~$of length $m_{1}h$,
$p(t)=p_{1}$, then for $t\in I_{2},~p(t)=p_{2}$ and so on until the $N$-th
time subinterval of length $m_{N}^{{}}h$ where $p(t)=p_{N}.$ Next, the
algorithm repeats. Relation (\ref{scheme}) is periodic with $T=(m_{1}%
+m_{2}+...+m_{N})h$. In order to simplify the notation, for a fixed step size
$h$, scheme (\ref{scheme}) will be denoted hereafter%

\begin{equation}
\lbrack m_{1}p_{1},~m_{2}p_{2},\ldots,m_{N}~p_{N}]. \label{schema simpla}%
\end{equation}

\noindent For example, for $N=3,$ by the scheme $\left[  1p_{1},3p_{2}%
,2p_{3}\right]  $ one should understand the infinite sequence of $p:$
$p_{1},p_{2},p_{2},p_{2},p_{3},p_{3},p_{1},\ldots~$which means that while
(\ref{eq1}) is integrated, $~p$ switches in each $I$ subinterval between the
values of $\mathcal{P}_{3}=\{p_{1},p_{2},p_{3}\}.$

\begin{remark}
In practical examples, switching techniques can be applied not only to
parameters but, for example, to the state variables \cite{Oksasoglua}.
\end{remark}

In order to compare two attractors, we introduce the following criteria

\begin{definition}
Two attractors will be considered approximately identical (AI) if after
neglected transients, their trajectories in the phase state are close enough
to each other.
\end{definition}

The AI property is understood as a perfect as possible overlap between orbits,
histograms and Poincar\'{e} sections (or, ideally, the same - perfect match).

It can be easy to verify the following property

\begin{proposition}
\textbf{\label{prop}~}\emph{For every }$N$\emph{, and }$\mathcal{P}_{N}%
,~$\emph{the relation (\ref{eq2}) can be written in the following form}
\end{proposition}

\begin{equation}
p^{\ast}=\frac{\sum\limits_{k=1}^{N}p_{k}m_{k}}{\sum\limits_{k=1}^{N}m_{k}}.
\label{p*}%
\end{equation}

\noindent\emph{Moreover, }$p^{\ast}$\emph{ is a convex combination of the
elements of }$\mathcal{P}_{N}$.

The last statement can be easy verified if we denote $\alpha_{k}=m_{k}%
/\sum\limits_{k=1}^{N}m_{k}.$ Next,~ because $\ \sum\limits_{k=1}^{N}%
\alpha_{k}=1,~p^{\ast}$ can be written: $p^{\ast}=\sum\limits_{k=1}^{N}%
\alpha_{k}p_{k}.$

\begin{notation}
Let denote by $A^{\ast}$ the \emph{synthesized attractor}, obtained with the
SA implemented by (\ref{schema simpla}) and by $A_{p^{\ast}}$ the
\emph{averaged attractor} obtained for $p=p^{\ast}$.
\end{notation}

Now, the property mentioned at the beginning of this section can be formulated
as follows:

\emph{For any }$N$\emph{ and }$P_{N}$\emph{, the synthesized attractor
}$A^{\ast}$\emph{ belongs within the set }$A_{N}.$

\smallskip The proof is presented in \cite{Yu}, but the result can be verified
by means of computational approach too. First, it can be verified
computationally that $A^{\ast}$ and $A_{p^{\ast}},~$with $p^{\ast}$given by
(\ref{p*}), are AI. Next, using Property \ref{prop}, $~p^{\ast}\in\left(
p_{1},p_{N}\right)  ,~$and taking into account the bijection between
$\mathcal{P}$ and $\mathcal{A}$, we are entitled to consider that the same
convex structure is preserved from $\mathcal{P}_{N}$ in $\mathcal{A}_{N}.$
Then $A_{p^{\ast}}\in\mathcal{A}_{N}~$\ and therefore $A^{\ast},$ which is AI
to $A_{p^{\ast}},$ belongs to $\left(  A_{1},A_{N}\right)  $ $.~$

\begin{remark}
i) The time subintervals $\Delta t_{i}$ and the size of the integration step
$h$~are parameters which may influence the results due to the convergence
properties of the considered method for ODEs. Therefore, after extensive
simulations, we have chosen the best possible values of $h$ so that the best
overlap is obtained$.$ However, $h$ is not a critical parameter. Therefore,
$\ $in almost all AS applications,$~$we chose usually values for $h$ (in this
paper $h=$ $0.005\div0.01).$\newline ii) To relatively large values for $m$ or
$N$ may correspond less or more significant differences between the two
attractors $A^{\ast}$ and $A_{p^{\ast}}~$(see \cite{Danca et al}), but
$A^{\ast}$ remains within of a relatively thin neighborhood of $A_{p^{\ast}}.$
\end{remark}

The pseudocode of periodic SA, applied on \ $I=[0,T_{\max}]$, for
chosen $N,~T_{\max},~h,~m_{1},\ldots,m_{N},~p_{_{1}},\ldots
p_{N},~$is presented in Table \ref{tab1}.

Due to the mentioned convex property, the scheme (\ref{schema
simpla}) may be applied in any random way \cite{Danca}. Therefore
the \emph{random }SA generates again an attractor $A^{\ast}$ which,
based on the mentioned above convexity property, will obviously
belong inside the set of considered attractors $\mathcal{A}_{N}$
endowed with the order of $\mathcal{P}_{N}$. The pseudo-code of one
of the possible variants is presented in Table \ref{tab2} where
$rand$ means some random generator (in this paper with uniform
distribution) of positive integers less than or equal to $N$.

\noindent $m_{i}^{\prime}$ count $p_{i}.$ $p^{\ast}$ is determined
with the following formula

\begin{equation}
p^{\ast}=\frac{\sum\limits_{k=1}^{N}p_{k}m_{k}^{\prime}}{\sum\limits_{k=1}%
^{N}m_{k}^{\prime}}. \label{rand number}%
\end{equation}

In this case, in order to obtain a better AI, the integration steps number
should be taken as large as possible.

\begin{remark}
\label{remark}SA cannot be considered as a "true" control algorithm (see
Section 3) even it may generate any stable trajectory for a considered system,
since before the algorithm starts, the system may evolve stable and then the
algorithm just changes the behavior from a stable attractor to another one.
The algorithm can be use as chaotification algorithm too, but again it should
not be considered as a real anticontrol algorithm (further informations on
chaos control can be obtained e.g. from \cite{Ogorzalek},\cite{Grebogi} and
for anticontrol of discrete and continuous dynamical systems \cite{Shi} and
\cite{Wang} respectively). The only condition for both control and anticontrol
is that $\mathcal{P}_{N}$ contain values corresponding to chaotic and stable
attractors too. It should be notified that the SA can be useful when a desired
value for $p$ cannot be set directly.\newline
\end{remark}

To see how the SA must be implement in practice, let us consider the sets
$\mathcal{P}_{N}$ and $\mathcal{A}_{N}$ and suppose that certain targeted
value of $\widehat{p}\notin$ $\mathcal{P}_{N}~$cannot be accessible, but we
want to generate the underlying attractor. By using the bifurcation diagram
for the considered dynamical system, the only sufficient condition on
$\widehat{p}$ is to belong to the real interval $(p_{1},p_{N})~$($\widehat{p}$
cannot be chosen outside this interval because of the mentioned convexity
property).$~$In order to synthesize the attractor $A_{\widehat{p}}$, we must
choose $m_{i}$ so that the desired value $\widehat{p}$ is given by (\ref{p*}).
This implies to solve (\ref{p*}) considered as an equation for fixed
$p_{1},\ldots,p_{N},$ with $p^{\ast}=\widehat{p},~$and unknowns $m_{i}$. With
the obtained $m_{i}~$values, scheme (\ref{schema simpla}) can next be applied.
The synthesized attractor $A^{\ast}$ will be identical, as shown above, to
$A_{\widehat{p}}$. Thus, by using the SA, one can "force" the system to evolve
on the desired trajectory corresponding to $\widehat{p}.$

Another practical situation is also possible: $\mathcal{P}_{N}$ and $m$ are
not known a priori.~Thus, $m$ and the set $\mathcal{P}_{N},$ have to be
determined so that relation (\ref{p*}) be verified with known $\widehat{p}$.

In both cases, the solutions are not unique because the elements of
$\mathcal{P}_{N}$ belong in a compulsory way to one of the infinite number of
$p$-intervals which may compose $\mathcal{P}$.

For example, let us consider the Lorenz system with the control parameter $p$,
and suppose we want to synthesize, with the scheme $[m_{1}p_{2},m_{2}p_{1}%
]~$for fixed $h,$ a stable trajectory corresponding to $\widehat{p}=150$
starting from $\mathcal{P}_{N}=\{130,190\}.~$Then, one of the possible
solutions to (\ref{p*}) is $m_{1}=2$ and $m_{2}=1$ for which (\ref{p*}) is
verified: $p^{\ast}=150=(2\ast130+1\ast190)/(2+1).$

\section{Finding the Rikitake's attractors}

\smallskip For the Rikitake system modeled by the equations (\ref{IVP}), we have%

\[
f(x)=\left(
\begin{array}
[c]{c}%
-ax_{1}+x_{2}x_{3}\\
-ax_{2}+x_{1}x_{3}\\
1-x_{1}x_{2}%
\end{array}
\right)  ,\text{ }A=\left(
\begin{array}
[c]{ccc}%
0 & 0 & 0\\
-1 & 0 & 0\\
0 & 0 & 0
\end{array}
\right)  .
\]

Throughout this paper the parameter $a$ is set to the value $a=1.$

In order to apply SA, the bifurcation diagram (Fig.\ref{fig4}) is a
useful tool to study the character of the attractors.

In the same figure, the attractors used in achieving the synthesis are plotted
together with the synthesized attractors.

\subsection{'Transient attractor'}

Numerical simulations of the Rikitake system suggest that the system has
attractors which are obviously bounded. In other words, the solutions enter a
ball around the origin from which they never escape. However, this seems not
to be true, since if we take as an initial condition $(0,0,k)~$the exact
solution of (\ref{IVP}) is $x_{1}(t)=x_{2}(t)=0,$ $x_{3}(t)=t+k$.
$\ $\smallskip But this solution is unstable, the $x_{3}$-axis being an
invariant manifold. There are orbits which escape to, or come from, infinity,
instead of going towards the attractor. If the flow is on the $x_{3}$-axis it
never escapes; if the flow is not on the $x_{3}$-axis, then it can never
enter. Because of the mentioned symmetry of solutions, for points arbitrary
close to the $x_{3}$-axis, the flow takes the trajectory back to a bounded
attractor (results on boundedness of solutions to third-order nonlinear
differential equations can be found in \cite{C.}).

Even though for the above mentioned reasons in some papers the
$x_{3}$-axis is regarded as unimportant to the dynamics of the
system, we found interesting dynamics due to $x_{3}$-axis
instability. Thus, for $p=90,~\ $we have found a new and intriguing
case which can be considered a kind of "transient attractor" TA (see
Fig.\ref{fig5} (a) where the three-dimensional plot is shown and
Fig.\ref{fig5} (d-f) where the phase projections have been drawn)$.$
TA is actually only a kind of extremely long time transient (its
existence being for $t\leq t^{\prime}~$with $t^{\prime}$ close to
$6\times10^{6}$) before the trajectory reaches the 'real' attractor,
a stable limit cycle denoted $L~~$(see Fig.\ref{fig5} (b) where one
can see the limit cycle $L$ whose magnified three-dimensional phase
plot is shown in Fig. 5 (c)). The dashed lines in Fig.\ref{fig5} (f)
indicate the well-defined movement sense of TA. The periodic
characteristics of $L$ can be observed in the three time series,
corresponding to $x_{1},$ $x_{2}$ and $x_{3}~$depicted in
Fig.\ref{fig5} (g), (j) and (m) for $t\in\left[ 0,10^{6}\right] $.
The dashed line indicates the moment $t=t^{\prime},~$when $L~$is
born. The details, $D_{1},~$for $t\leq t^{\prime}:$ $~t\in\left[
9.65\times 10^{5},9.75\times10^{5}\right] $ are presented in
Fig.\ref{fig5} (h), (k) and (n).
For $t>t^{\prime},$ the TA transforms into $L~.~$Magnified details, $D_{2}%
,~$for $t\in\left[  7\times10^{6},7.1\times10^{6}\right]  ~$are
presented in Fig.\ref{fig5} (i),(l),(o))$.~$Before TA ends in $L$,
it crosses it several times. It can be seen that the size of TA is
of the order of $10^{3}$ as compared to the small size of the $L$
which is nearly $1000$ times smaller. Because a part of the
trajectory lies on the $x_{3}$-axis (or is very close to it (see
e.g. Fig.\ref{fig5} (d),(e),(f)), the distance being in this case
not highlighted by the numerical method), we suspect that the TA
appears because of $x_{3}$-axis instability. On this path along the
$x_{3}$-axis, the speed of the TA is very slow as compared to the
loop speed (see the vertical peeks in the time series in
Fig.\ref{fig5} (g),(j),(m) and the details $D_{1}~$in Fig.\ref{fig5}
(h),(k),(n)). Being a stiff system, not all numerical methods have
proved to be adequate to obtain the TA. However, the many
simulations and the TA geometric symmetry, encouraged us to consider
TA as being not just some "false" trajectory due to numerical
integration, but a real representative component of the insight
dynamics of the Rikitake system.

\subsection{Attractors synthesis}

By applying the deterministic or random SA, any attractor of Rikitake's system
may be synthesized. The standard Runge-Kutta method has been utilized and the
most representative cases has been considered. Thus, $\mathcal{P}_{N}$ has
been chosen so that all kind of behaviors (regular and chaotic) may be considered.

\begin{figure}
\begin{center}
  \includegraphics[clip,width=0.7\textwidth]{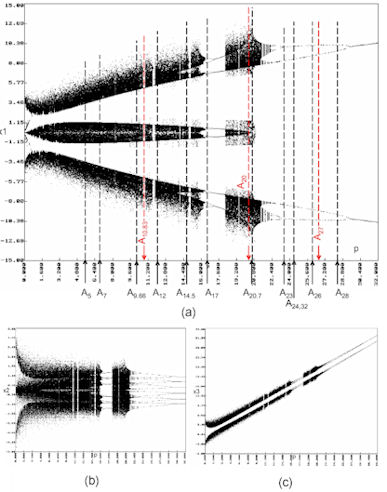}
  \caption{Bifurcation diagram for the Rikitake system. (a) Bifurcation
diagram for the component $x_{1}$ where the positions of the
attractors utilized and the synthesized attractors are indicated;
(b), (c) Bifurcation diagram for components $x_{2}$ and $x_{3}$.}
\label{fig4}
\end{center}
\end{figure}

\begin{figure}
\begin{center}
  \includegraphics[clip,width=0.7\textwidth]{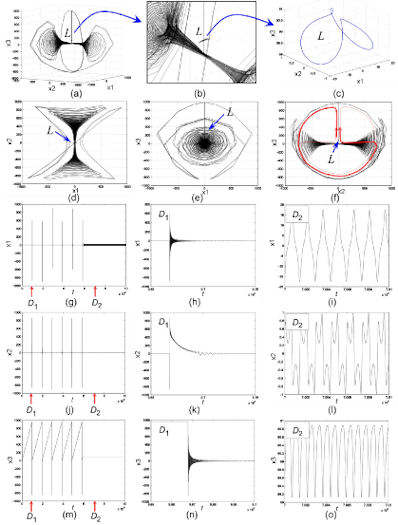}
  \caption{Transient attractor (TA) and stable limit cycle L. (a) Phase plot; (b)
detail; (c) Limit cycle L after neglected TA; (d), (e), (f)
projections of TA and L; (g) Time series for $x_{1}~$component; (h),
(i) the details $D_{1}$ and $D_{2}$ indicated in (g); (j) Time
series for component $x_{2}$; (k),(l) details $D_{1}$ and $D_{2}$
indicated in (j); (m) Time series for component $x_{3}$; (n),(o)
Details $D_{1}$ and $D_{2}$ indicated in (m).}\label{fig5}
\end{center}
\end{figure}

\begin{figure}
\begin{center}
  \includegraphics[clip,width=0.7\textwidth]{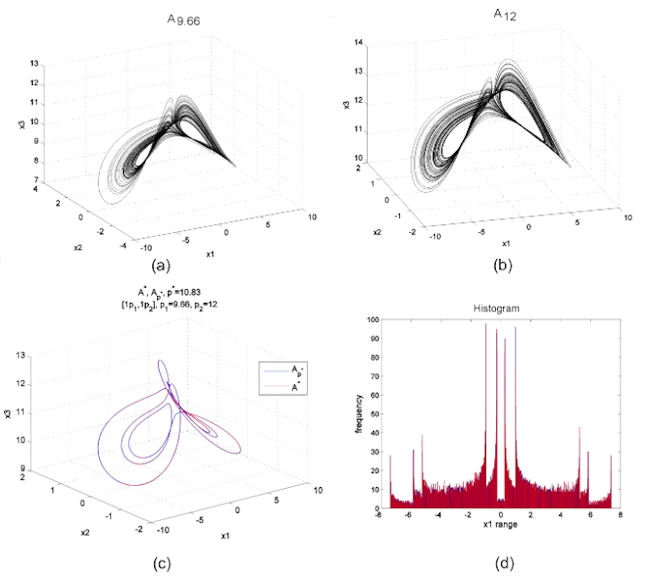}
  \caption{ Synthesized attractor $A^{\ast}$ and $A_{p^{\ast}}$ obtained with
scheme $[1p_{1},1p_{2}]$ with $p_{1}=9.66$ and $p_{2}=12$;
$p^{\ast}=10.83.$ (a), (b) Phase plots of $A_{p_{1}}$ and
$A_{p_{2}};~$(c) $A^{\ast}$ and $A_{p^{\ast}}$ superimposed; (d)
Histograms superimposed.}\label{fig6}
\end{center}
\end{figure}

Using the deterministic scheme $[1p_{1},1p_{2}]$ with $p_{1}=9.66$
and $p_{2}=12$, one obtains the synthesized attractor
$A^{\ast}$(Fig.\ref{fig6} (c))$.$ Fig.\ref{fig6} (a) and (b) present
$A_{p_{1}}$ and $A_{p_{2}}$. $A^{\ast}$ is identical with
$A_{p^{\ast}}$ for $p^{\ast}$ given by (\ref{p*}) $p^{\ast }=\left(
p_{1}+p_{2}\right) /2=10.83$ (see Fig.\ref{fig6} (c) where both
$A^{\ast}$ and$~A_{p^{\ast}}$ are plotted superimposed and
Fig.\ref{fig6} (d) where histograms of both attractors are
presented). It should be noticed that, in this case, SA can be
viewed as control algorithm (see Remark \ref{remark} i) since
$A_{p_{1}}$ and $A_{p_{2}}$ are chaotic and $A^{\ast}$ is a stable
limit cycle (see Fig.\ref{fig4}).

\begin{figure}
\begin{center}
  \includegraphics[clip,width=0.7\textwidth]{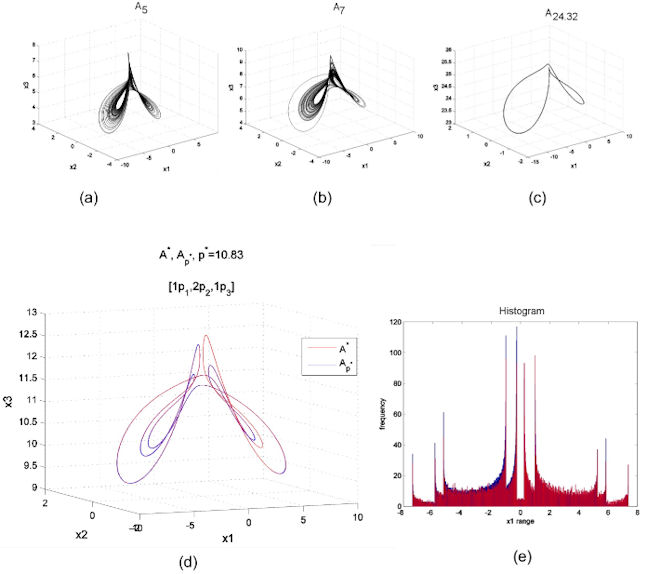}
  \caption{Synthesized attractor $A^{\ast}$ and $A_{p^{\ast}}$ obtained with
scheme $[1p_{1},2p_{2},1p_{3}]$ with $p_{1}=5,$
$p_{2}=7,~p_{3}=24.32$; $p^{\ast}=10.83.$ (a)-(c) Phase plots of
$A_{p_{1,2,3}}.$ (d) $A^{\ast}$ and $A_{p^{\ast}}$ superimposed; (e)
Histograms superimposed.}\label{fig7}
\end{center}
\end{figure}

An attractor can be obtained within several variants of (\ref{schema simpla}).
For example $A_{10.83}$ synthesized bellow, can be obtained too with the
scheme $[1p_{1},2p_{2},1p_{3}]$ with $p_{1}=5,$ $p_{2}=7$ and $p_{3}%
=24.32~$(see Fig.\ref{fig4} and Fig.\ref{fig7}). Here, $p^{\ast}=10.83=(p_{1}+2p_{2}%
+p_{3})/4.$ $A^{\ast}$ and $A_{p^{\ast}}$ are plotted superimposed
in Fig.\ref{fig7} (d). Superimposed histograms (Fig.\ref{fig7} (e))
underline the identity.

\begin{figure}
\begin{center}
  \includegraphics[clip,width=0.7\textwidth]{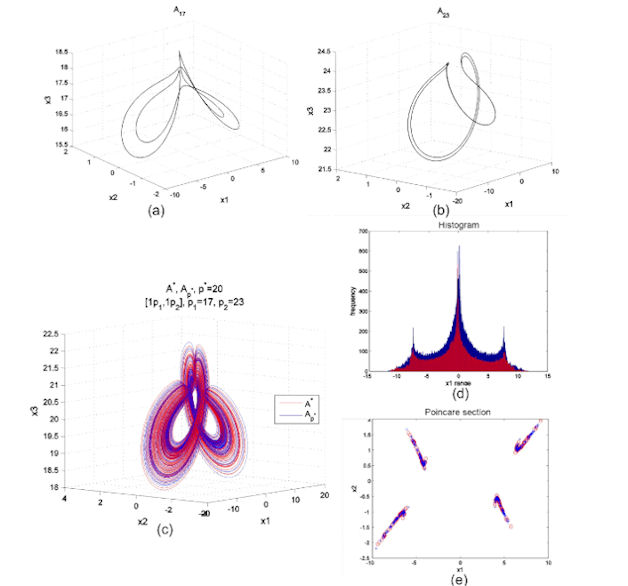}
  \caption{Synthesized attractor $A^{\ast}$ and $A_{p^{\ast}}$ obtained with
scheme $[1p_{1},1p_{2}]$ with $p_{1}=17,$ $p_{2}=23$;
$p^{\ast}=20.~$ (a), (b) Phase plots of $A_{p_{1}}$ and
$A_{p_{2}};~$(c) $A^{\ast}$ and $A_{p^{\ast}}$ superimposed; (d)
Histograms superimposed; (e) Poincar\'{e} sections superimposed.}
\label{fig8}
\end{center}
\end{figure}

With the scheme $[1p_{1},1p_{2}]$ with $p_{1}=17~$and $p_{2}=23$ the
chaotic attractor $A^{\ast},$ which is identical to $A_{p^{\ast}}$
with $p^{\ast}=20$ (Fig.\ref{fig4} and Fig.\ref{fig8} (c)), can be
obtained. Superimposed histograms and the Poincar\'{e} section
(Fig.\ref{fig8} (d),(e)) demonstrate the identity. Here SA could be
considered as an anticontrol algorithm (see Remark \ref{remark} i).

\begin{figure}
\begin{center}
  \includegraphics[clip,width=0.7\textwidth]{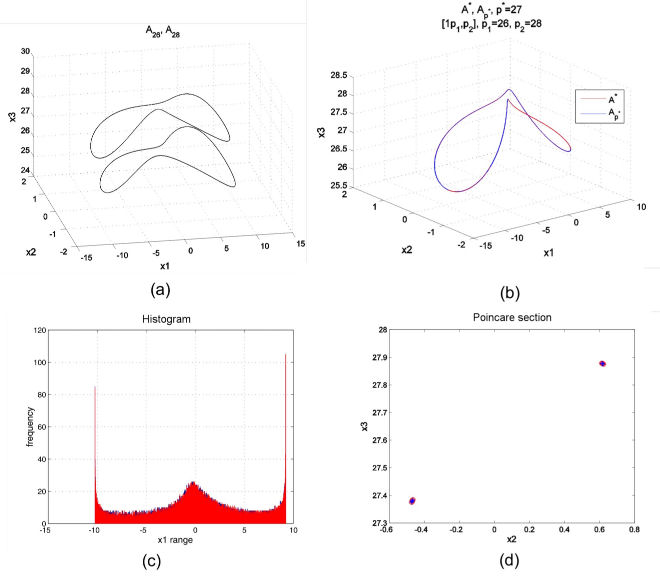}
  \caption{Synthesized attractor $A^{\ast}$ and $A_{p^{\ast}}$
obtained with scheme $[1p_{1},1p_{2}]$ with $p_{1}=26,$ $p_{2}=28$;
$p^{\ast}=27.~$(a) Phase plots of $A_{p_{1}}$ and $A_{p_{2}}$; (b)
$A^{\ast}$ and $A_{p^{\ast}}$ superimposed; (c) Histograms
superimposed; (d) Poincar\'{e} sections superimposed.}\label{fig9}
\end{center}
\end{figure}

If between $p_{1}$ and $p_{2},$ corresponding to stable attractors,
there are no chaotic windows, anticontrol cannot be realized (Remark
\ref{remark} i). For example, by choosing $p_{1}=26$ and $p_{2}=28$
for which $A_{p_{1}}$ and $A_{p_{2}}$ are stable limit cycles (see
Fig.\ref{fig4} and Fig.\ref{fig9} (a) where both
attractors are plotted together), whatever kind of scheme (\ref{schema simpla}%
) is used, only stable limit cycle can be obtained. For example,
with the $p_{1,2\text{ }}$chosen bellow, and the scheme
$[1p_{1},2p_{2}]~$one obtains the limit cycle depicted in
Fig.\ref{fig9} (b). The identity between $A^{\ast}$ and
$A_{p^{\ast}}$ with $p^{\ast}=27$ is underlined by histograms and
Poincar\'{e} sections superimposed in Fig.\ref{fig9}(d) and (e).

\begin{figure}
\begin{center}
  \includegraphics[clip,width=0.7\textwidth]{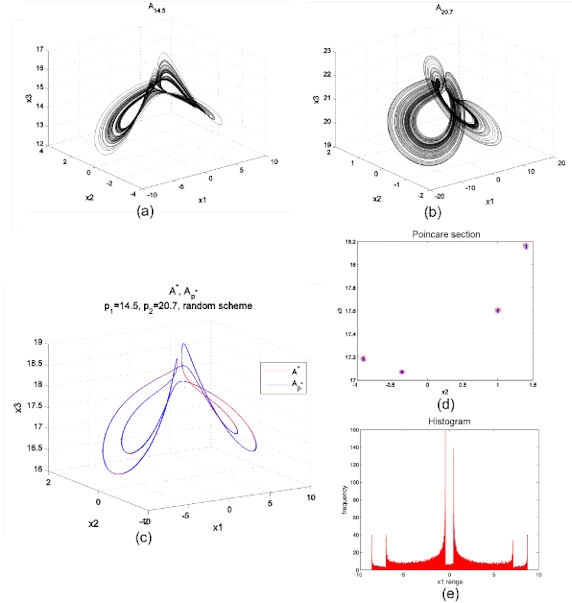}
  \caption{Synthesized attractor $A^{\ast}$ and $A_{p^{\ast}}$ obtained with
random scheme with $p_{1}=14.5,$ $p_{2}=20.7;$ $p^{\ast}=17.6.$ (a),
(b) Phase plots of $A_{p_{1}}$ and $A_{p_{2}};~$(c) $A^{\ast}$ and
$A_{p^{\ast}}$ superimposed; (d) Poincar\'{e} sections superimposed;
(e) Histograms superimposed.}\label{fig10}
\end{center}
\end{figure}

As stated in Section 2, attractors can be synthesized by using
random SA. Thus, by choosing the uniform random distribution with
algorithm presented in Table \ref{tab2}, for $p_{1}=14.5$ and
$p_{2}=20.7,$ one obtains $A^{\ast}$ identical with $A_{p^{\ast}}$
with $p^{\ast}=17.6~\ $given by (\ref{rand number}) (Fig.\ref{fig4}
and Fig.\ref{fig10}(c)).

The results are presented in Table \ref{tab3}.

\section{\smallskip Conclusions and further directions}

In this paper we verified numerically and computationally that any attractor
of the Rikitake system can be obtained with the SA, following some
deterministic or random rule, the analytical proof being presented in
\cite{Yu}.

In addition to the fact that the SA may force the system to evolve along any
attractor, it introduces a convex structure inside the attractors set.

Also, SA may serve as a model to explain the born of specific dynamics in
systems encountered in the real world or in experiments where this kind of
deterministic or random parameter switches may occur.

In this study we extended the application of the SA from systems with
hyperbolic equilibrium to systems with non-hyperbolic ones. We have also found
an interesting transient attractor, TA, which persists for a long period of
time. Its presence is supposed to be due to the instability along the $x_{3}%
$-axis, and not all numerical methods for ODEs can reveal it. Representing an
interesting component of Rikitake's dynamics, the TA worth to be studied
rigorously, for example from the point of view of shadowing theory.

The rigorous proof of the existence of the bijection between the set of
parameter values and the set of attractors and the analytical proof of the
application of the SA to other class of systems represent tasks for future studies.

\smallskip\newpage

\begin{table}[ht]
\caption{Pseudo-code of the $\emph{PS}$ algorithm} \label{tab1}
\begin{center}
\begin{tabular}[c]{l}
\hline\noalign{\smallskip}
repeat\\
\quad~~\,for~$i=1$~to~$m_{1}$~do\\
~~~~~~~~~~~~~~~~~integrate~(\ref{eq1})~for~$p=p_{_{1}}$\\
~~~~~~~~~~~~~~~~~$t=t+h$\\
~~~~~end\\
~~~~~~~~~~~...\\
~~~~~\,for~$i=1$~to~$m_{N}$~do\\
~~~~~~~~~~~~~~~~integrate~(\ref{eq1})~for~$p=p_{_{N}}$\\
~~~~~~~~~~~~~~~~$t=t+h$\\
~~~~~end\\
until~$t\geq T_{\max}$
\end{tabular}
\end{center}
\end{table}

\begin{table}[ht]
\caption{Pseudo-code of the random $\emph{PS}$ algorithm}
\label{tab2}
\begin{center}
\begin{tabular}[c]{l}
\hline\noalign{\smallskip}
repeat\\
~~~~label=$\operatorname{rand}(N)$\\
~~~~if~$label=1$~then\\
~~\ ~~~~~~~integrate~(\ref{eq1})~with~$p=p_{_{1}}$\\
~\ \ \ \ \ \ \ \ \ $m_{1}^{\prime}=m_{1}^{\prime}+1$\\
~~~~if~$label=2$~then\\
~~~~~~~~~~integrate~(\ref{eq1})~with~$p=p_{_{2}}$\\
~\ \ \ \ \ \ \ \ \ $m_{2}^{\prime}=m_{2}^{\prime}+1$\\
~\ \ \ \ \vdots\\
~\ ~~if~$label=N$~then\\
~~~~~~~~~~integrate~(\ref{eq1})~with~$p=p_{N}$\\
~\ \ \ \ \ \ \ \ \ $m_{N}^{\prime}=m_{N}^{\prime}+1$\\
~~\ ~~$t=t+h$\\
until~$t\geq T_{\max}$
\end{tabular}
\end{center}
\end{table}

\begin{table}
\caption{The results of PS algoritm applied to the Rikitake system
(\ref{IVP})}\begin{center}\label{tab3}
\begin{tabular}
[c]{||c|c|c|c|c|c||}\hline\hline {\footnotesize Scheme} &
{\footnotesize p}$_{1}$ & {\footnotesize p}$_{2}$ &
{\footnotesize p}$_{3}$ & {\footnotesize p}$^{\ast}$ & {\footnotesize Remarks}%
\\\hline
{\footnotesize [1p}$_{1}${\footnotesize ,1p}$_{2}${\footnotesize ]}
& {\footnotesize 9.66} & {\footnotesize 12} & {\footnotesize -} &
{\footnotesize 10.83} & {\footnotesize A}$_{p_{1}}${\footnotesize ,~A}$_{p_{2}%
}${\footnotesize ~chaotic,~A}$^{\ast}${\footnotesize ~stable limit
cycle (Fig.\ref{fig6})}\\\cline{1-3}\cline{2-6}%
{\footnotesize [1p}$_{1}${\footnotesize ,2p}$_{2},${\footnotesize ,1p}$_{3}%
${\footnotesize ]} & {\footnotesize 5} & {\footnotesize 7} &
{\footnotesize 24.32} & {\footnotesize 10.83} & {\footnotesize A}$_{p_{1}}%
,${\footnotesize ~A}$_{p_{2}}~${\footnotesize chaotic, A}$_{p_{2}}%
${\footnotesize stable limit cycle,~A}$^{\ast}${\footnotesize
~stable limit
cycle (Fig.\ref{fig7})}\\\cline{1-3}\cline{2-6}%
{\footnotesize [1p}$_{1}${\footnotesize ,1p}$_{2}${\footnotesize ]}
& {\footnotesize 17} & {\footnotesize 23} & {\footnotesize -} &
{\footnotesize 20} & {\footnotesize A}$_{p_{1}},${\footnotesize ~A}$_{p_{2}}%
~${\footnotesize stable limit cycles, A}$^{\ast}${\footnotesize
~chaotic~(Fig.\ref{fig8})}\\\hline {\footnotesize
[1p}$_{1}${\footnotesize ,1p}$_{2}${\footnotesize ]} &
{\footnotesize 26} & {\footnotesize 28} & {\footnotesize -} &
{\footnotesize 27} & {\footnotesize A}$_{p_{1}}$, {\footnotesize ~A}$_{p_{2}}%
${\footnotesize stable limit cycles, A}$^{\ast}${\footnotesize
~stable limit cycle (Fig.\ref{fig9} )}\\\hline {\footnotesize random
SA} & {\footnotesize 14.5} & {\footnotesize 20.7} &
{\footnotesize -} & {\footnotesize 17.6} & {\footnotesize A}$_{p_{1}}%
${\footnotesize ,~A}$_{p_{2}} ${\footnotesize ~chaotic attractors, A}$^{\ast}%
${\footnotesize ~stable limit cycle (Fig.\ref{fig10})}\\\hline\hline
\end{tabular}
\end{center}
\end{table}

\clearpage

\end{document}